\documentclass[twocolumn,showpacs,preprintnumbers,amsmath,amssymb]{revtex4}

\usepackage{graphicx}% Include figure files
\usepackage{dcolumn}% Align table columns on decimal point
\usepackage{bm}% bold math
\begin{document}

\title{Nuclear transparencies from photoninduced pion production}

\author{W. Cosyn}
\email{Wim.Cosyn@UGent.be}
\author{M. C. Mart\'{\i}nez}
\altaffiliation{Present address: Dpto FAMN, Universidad Complutense de 
Madrid, E-28040 Madrid, Spain.}
\author{J. Ryckebusch}
\author{B. Van Overmeire}

\affiliation{Department of Subatomic and Radiation Physics,\\
 Ghent University, Proeftuinstraat 86, B-9000 Gent, Belgium}
\date{\today}

\begin{abstract}
We present a relativistic and cross-section factorized framework for
computing nuclear transparencies extracted from $A(\gamma,\pi N)$
reactions at intermediate energies.  The proposed quantummechanical
model adopts a relativistic extension to the multiple-scattering
Glauber approximation to account for the final state interactions of
the ejected nucleon and pion.  The theoretical predictions are compared to the
experimental $^4$He$(\gamma,p \pi^-)$ data from Jefferson Lab.  For
those data, our results show no conclusive evidence for the onset of
mechanisms
related to color transparency.
\end{abstract}


\maketitle 

The strong force exhibits a strong scale dependence. At low energies,
hadrons are undoubtedly the adequate degrees of freedom. The
properties of nuclei can be fairly well understood in a picture in
which nucleons exchange mesons. At high energies, that particular role
of fermions interacting through force-carrying bosons, is played by
quarks and gluons. The transition energy region is a topic of current
intensive research, for example at Jefferson Lab, where hadronic
matter can be studied with intense beams of real and virtual photons
possessing the proper wavelengths in the femtometer and sub-femtometer
range.  A commonly used observable to pin down the underlying dynamics
of hadronic matter is the nuclear transparency to the transmission of
hadrons. The nuclear transparency for a certain reaction process is
defined as the ratio of the cross section per target nucleon to the
one for a free nucleon. Accordingly, the transparency is a measure for
the effect of the medium on the passage of energetic hadrons.  It
provides an excellent tool to search for deviations from predictions
of models based on traditional nuclear physics.  One such phenomenon 
is color transparency (CT).  Color transparency predicts the reduction
of final state interactions (FSI) of hadrons propagating through
nuclear matter in processes at high momentum transfer.  Experiments
have been carried out to measure nuclear transparencies in search of
CT in $A(p,2p)$
\cite{Carroll:1988rp,Mardor:1998,Leksanov:2001ui,Aclander:2004zm} and
$A(e,e'p)$ \cite{Garino:1992ca,Makins:1994mm,O'Neill:1994mg,
Abbott:1997bc,Garrow:2001di,Dutta:2003yt} reactions, $\rho$-meson
production \cite{Adams:1994bw,Airapetian:2002eh} and diffractive
dissociation of pions into di-jets \cite{Aitala:2000hc}.  
%The $A(p,2p)$ experiments show an increase followed by a
%decrease of the transparency for $Q^2 \approx 3-10~(\text{GeV/c})^2$.
%The transparencies extracted from $A(e,e'p)$ experiments can be
%reproduced in calculations with traditional nuclear models up to $Q^2
%= 8.1~ (\text{GeV/c})^2$.  Results on vector-meson production reported
%in Refs.~ \cite{Adams:1994bw,Airapetian:2002eh,Aitala:2000hc} are in
%agreement with calculations including CT. 
Intuitively one expects CT
to reveal itself more rapidly in reactions involving mesons. Indeed,
it appears more probable to produce a two-quark configuration with a
small transverse size and, as a consequence, reduced FSI.

Recently, the nuclear transparency for the pion photoproduction
process $\gamma n \rightarrow \pi^- p$ in $^4\text{He}$ has been
measured in Hall A at Jefferson Lab \cite{Dutta:2003mk}.  In the
experiment, both the outgoing proton and the pion are detected for
photon beam energies in the range 1.6-4.5 GeV and pion center-of-mass
scattering angles of $70^o$ and $90^o$.  With two hadrons in the final
state, the pion photoproduction reaction on helium offers good
opportunities to search for medium effects that go beyond traditional
nuclear physics.  Not only is He a dense system, its small radius
optimizes the ratio of the hadron formation length to the size of the
system.  Along the same line, a pion electroproduction experiment at
Jefferson Lab, measuring pion transparencies in H, D, $^{12}\text{C}$,
$^{64} \text{Cu}$ and $^{197} \text{Au}$ \cite{Garrow:2001fk} has been
completed and data analysis is currently under way.  It speaks for
itself, though, that the availability of a model for computing
transparencies extracted from $A(\gamma,\pi N)$ and $A(e,e' \pi N)$,
is essential for interpreting all these measurements. As a matter of
fact, it is a real challenge to compute the effect of the medium on
the simultaneous emission of a nucleon and a pion within the context
of traditional nuclear-physics models.

In Ref.~\cite{Dutta:2003mk} the results for $\gamma n \rightarrow
\pi^- p$ on $^4\text{He}$ are compared to the predictions of a
semi-classical model by Gao, Holt and Pandharipande \cite{Gao:1996mg}.
The results of the calculations including CT effects were found to be
more consistent with the measurements.  An alternative semi-classical
approximation to compute pionic transparencies is presented in
\cite{Larson:2006ge}.  A quantummechanical model for meson
photoproduction on $^2$H was recently proposed by J.-M. Laget
\cite{Laget:2006bu}. In this Letter, we present a quantummechanical
model for computing the nuclear transparencies in $ \gamma n
\longrightarrow \pi ^{-} p$ on finite nuclei. The model is based on a
relativized version of Glauber theory and is essentially
parameter-free. To our knowledge the presented framework is the first
of its kind.

In describing the $A(\gamma,N \pi) A-1 $ reaction we use the
following lab four-momenta: $Q^\mu (q,\vec{q})$ for the photon,
$P^\mu_A (E_A,\vec{p}_A= \vec{O})$ for the target nucleus, $P^\mu_{A-1}
(E_{A-1},\vec{p}_{A-1})$ for the residual nucleus, $P^\mu_N =
(E_N,\vec{p}_N)$ and $P^\mu_\pi = (E_\pi,\vec{p}_\pi)$ for the ejected
nucleon and pion.  The missing momentum $\vec{p}_m$ is defined as
$\vec{p}_m= -\vec{p}_{A-1} =\vec{p}_N + \vec{p}_\pi -
\vec{q}$. Following the conventions of Ref.~\cite{Bjorken:Drell}, the
fivefold differential cross section reads in the lab frame 
\begin{equation}
\label{cross}
\frac{d \sigma}{dE_\pi d\Omega_\pi d\Omega_N} = \frac{M_{A-1} m_N
p_\pi p_N}{4(2\pi)^5q E_A} f^{-1}_{rec} \overline{\sum_{if}}
|\mathcal{M}_{fi}^{(\gamma,N\pi)}|^2\,,
\end{equation}
with the recoil factor given by:
\begin{equation}
f_{rec}=\frac{E_{A-1}}{E_A}\left|
1+\frac{E_N}{E_{A-1}}\left(1+\frac{(\vec{p}_\pi - \vec{q})\cdot
\vec{p}_N }{p^2_N}\right)\right|\,, 
\end{equation}
and 
$\mathcal{M}_{fi}^{(\gamma,N\pi)}$ the invariant matrix element:
\begin{equation} \label{matrixel}
\mathcal{M}_{fi}^{(\gamma,N\pi)} = \langle P_\pi^{\mu}, P_N^{\mu} m_s,
P_{A-1}^{\mu} J_R M_R | \hat{\mathcal{O}} | Q^{\mu}, P_A^{\mu} 0^+
\rangle \, ,
\end{equation}
where $m_s$ is the spin of the ejected nucleon $N$ and $J_R M_R$ the
quantum numbers of the residual nucleus.  We restrict ourselves to
processes with an even-even target nucleus $A$.  The operator
$\hat{\mathcal{O}}$ describes the pion photoproduction process and we
assume it to be free from medium effects. This is a common assumption
in nuclear and hadronic physics and is usually referred to as the
impulse approximation (IA).

For the target and residual nucleus we use relativistic wave functions
as they are obtained in the Hartree approximation to the $\sigma
\omega$-model with the W1 parameterization \cite{Furnstahl:1996wv}.
When studying the transparency, it is convenient to factorize the
invariant matrix element $\mathcal{M}_{fi}^{(\gamma,N\pi)}$ into a
part containing the elementary pion photoproduction process and a
part with the typical medium mechanisms in the process under study.
It is clear that the attenuation on the ejected proton and pion
induced by FSI mechanisms belongs to the last category and determines
the nuclear transparency for the process under study.  Even in the
relativistic plane-wave limit for the ejected nucleon and pion wave
function, factorization of the cross section is not reached through
the presence of negative-energy contributions.  Neglecting these, the
computation leads to an expression for the cross section in the
relativistic plane wave impulse approximation (RPWIA)
\begin{eqnarray}
\left( \frac {d \sigma} 
{dE_\pi d\Omega_\pi d\Omega_N} \right)_ {\text{RPWIA}} \approx 
\nonumber \\ 
\frac{ M_{A-1} p_\pi p_N \biggl( s-\left( m_{N} \right)^2 \biggr)^2} 
{4 \pi m_{N} q M_A} f_{\text{rec}}^{-1}
\rho ^ {\alpha} (\vec{p}_m) \frac {d\sigma^{\gamma\pi}} {d \mid t \mid}  \,,
\label{crossnoFSI}
\end{eqnarray}
with $\rho ^{\alpha} (\vec{p}_m)=
\sum_{m_s,m}|\bar{u}(\vec{p}_m,m_s)\phi_{\alpha}(\vec{p}_m)|^2$ the
momentum distribution which is obtained by contracting the bound-state
wave function $\phi_{\alpha}$ with the Dirac spinor
$\overline{u}$. The $\alpha (n,\kappa,m)$ denotes the quantum numbers
of the bound nucleon on which the photon is absorbed. Further, $ \frac
{d \sigma^{\gamma\pi}} {d\mid t \mid} $
denotes the cross section for $\gamma + N \longrightarrow \pi + N' $,
and $s=\left( Q ^{\mu} + P_A^{\mu} \right) ^2$ and $t= \left( Q^{\mu} -
P_{\pi} ^{\mu} \right) ^2$ are the Mandelstam variables
%, and $m_{N}^{*}$ stands for the average mass of a bound nucleon
.

In this work, we concentrate on $A(\gamma,N \pi) A-1 $ processes for
which the wavelengths of the ejected nucleons and pions
 are typically smaller than their interaction
ranges with the nucleons in the rest nucleus. Those conditions
make it possible to describe the FSI mechanisms with the aid of a
Glauber model. A relativistic extension of the Glauber model, dubbed
the Relativistic Multiple-Scattering Glauber Approximation (RMSGA),
was introduced in Ref.~\cite{Ryckebusch:2003fc}. In the RMSGA, the
wave function for the ejected nucleon and pion is a convolution of a
relativistic plane wave and an eikonal Glauber phase operator
$\widehat{\mathcal{S}}_{\text{FSI}}(\vec{r})$ which accounts for all FSI
mechanisms. Through the operation of
$\widehat{\mathcal{S}}_{\text{FSI}}(\vec{r})$ every residual nucleon in the
forward path of the outgoing pion and nucleon adds an extra phase to
their wave function.  The RMSGA framework has proved succesful in
describing cross sections and other observables in exclusive
$A(e,e'p)$ \cite{Ryckebusch:2003fc,Lava:2004zi} and $A(p,2p)$
\cite{VanOvermeire:2006dr} reactions.  The numerically challenging
component in RMSGA is that $\widehat{\mathcal{S}}_{\text{FSI}}(\vec{r})$
involves a multiple integral which tracks the effect of all collisions
of an energetic nucleon and pion with the remaining nucleons in the
target nucleus.  Realistic nuclear wave functions are also used in the models of
Refs.~\cite{Gao:1996mg,Larson:2006ge}. Contrary to the RMSGA model,
however, the transparencies are computed at the squared amplitude
level adopting a semi-classical picture for the FSI mechanisms.

In the numerical calculations within the context of the RMSGA, the
following phase is added to the product wave function for the ejected
nucleon and pion:
\begin{eqnarray}
\label{glauberphase}
\widehat{\mathcal{S}}_{\text{FSI}}(\vec{r}) =
%\widehat{\mathcal{S}}_{\text{FSI}}(\vec{r},\vec{r}_2,\ldots,\vec{r}_A)=
\prod_{j=2}^A
\left[ 1-\Gamma_{N'N} (\vec{b}-\vec{b}_j)\theta(z_j-z)\right]
\nonumber \\
\times 
 \left[ 1-\Gamma_{\pi N} ( \vec{b}' -\vec{b}'_j
) \theta(z'_j - z') \right]\,,
\end{eqnarray}
where $\vec{r}_j (\vec{b}_j, z_j) $ are the coordinates of the
residual nucleons and $\vec{r} (\vec{b},z) $ specifies the interaction
point with the photon.  In Eq. (\ref{glauberphase}), the $z$ and $z'$
axis lie along the path of the ejected nucleon and pion respectively. The
$\vec{b}$ and $\vec{b}'$ are perpendicular to these paths.
Reflecting the diffractive nature of the nucleon-nucleon ($N'N$) and
pion-nucleon ($\pi N $) collisions at intermediate energies, the
profile functions $\Gamma_{N'N}$ and $\Gamma_{\pi N}$ in
Eq. (\ref{glauberphase}) are parameterized as
\begin{equation}
\label{eq:profile}
\Gamma_{iN} (\vec{b}) =
\frac{\sigma^{\text{tot}}_{iN}(1-i\epsilon_{iN})}
{4\pi\beta_{iN}^2}\exp \left( {-\frac{\vec{b}^2}{2\beta_{iN}^2}}
\right) \,\;
(\textrm{with,} \; i = \pi \; \textrm{or} \; N') \; .
\end{equation}
Here, the parameters $\sigma^{\text{tot}}_{iN}$ (total cross section),
$\beta_{iN}$ (slope parameter) and $\epsilon_{iN}$ (real to imaginary
part ratio of the amplitude) depend on the momentum of the outgoing
particle $i$. In our calculations those parameters are obtained by
interpolating data from the databases for $ N' N \longrightarrow N' N$
from the Particle Data Group \cite{PDBook} and $ \pi N \longrightarrow
\pi N$ from the analysis of Refs.~\cite{Arndt:2003if,Workman}.

Now we derive an expression for the fivefold $A(\gamma,N \pi) A-1 $
cross sections when implementing FSI effects. To this end, we define
the distorted momentum distribution: 
\begin{equation}
\rho ^{\alpha} _{RMSGA}(\vec{p}_m)=
\sum_{m_s,m}|\bar{u}(\vec{p}_m,m_s) \phi^D_{\alpha}(\vec{p}_m)|^2 \; .
\label{eq:rhormsga}
\end{equation}
Here, $ \phi^D_{\alpha}(\vec{p})=\frac{1}{(2\pi)^{3/2}}\int d
\vec{r} e^{-i\vec{p}\cdot\vec{r}}\phi_{\alpha}(\vec{r})
\widehat{\mathcal{S}}_{\text{FSI}}^\dagger(\vec{r}) $ is the distorted
momentum-space wave function, which is the Fourier transform of the
bound nucleon wave function and the total Glauber phase.
In the absence of FSI, the $\rho_{RMSGA} ^{\alpha} (\vec{p}_m)$ of
Eq.~(\ref{eq:rhormsga}) reduces to the $\rho ^{\alpha} (\vec{p}_m)$ in
Eq.~(\ref{crossnoFSI}) when negative-energy components are neglected.
Based on this analogy we obtain the cross section in the RMSGA approach by replacing $\rho ^{\alpha} (\vec{p}_m) $ by $ \rho ^{\alpha}_{\text{RMSGA}}(\vec{p}_m) $ in Eq.~(\ref{crossnoFSI}).
% Based on this analogy we write the cross section in the RMSGA approach as 
% \begin{eqnarray}
% \label{crossFSI}
% \left(\frac{d  \sigma}{dE_\pi d\Omega_\pi
% d\Omega_N}\right)_{\text{RMSGA}} \approx \nonumber \\ 
% \frac{M_{A-1}
% p_\pi p_N \biggl( s- \left( m_{N} \right) ^2 \biggr)^2}
% {4\pi m_{N} qM_A} f_{\text{rec}}^{-1}
% \rho ^{\alpha} _{\text{RMSGA}}(\vec{p}_m) \frac{d\sigma^{\gamma\pi}}{d \mid t \mid} \, ,
% \end{eqnarray}
% which is obtained by replacing  the
% $\rho ^{\alpha} (\vec{p}_m) $ by $ \rho ^{\alpha}
% _{\text{RMSGA}}(\vec{p}_m) $ in Eq.(\ref{crossnoFSI}). 
%
%
% introduce color transparency
%

In our calculations, color transparency effects are implemented in the
standard fashion by replacing the total cross sections
$\sigma^{\text{tot}}_{iN}$ in Eq.~(\ref{eq:profile}) with effective ones
\cite{Farrar:1988me} which account for some reduced interaction over a
typical length scale $l_h$ corresponding with the hadron formation
length ($i = \pi \; \textrm{or} \; N'$)
\begin{equation}
\frac { \sigma^{\text{eff}}_{iN} }  
{ \sigma^{\text{tot}}_{iN} } =  \biggl\{ \biggl[
 \frac{\mathcal{Z}}{l_h} + \frac{<n^2 k_t^2>}{t} 
\left( 1-\frac{\mathcal{Z}}{l_h} \right) \theta(l_h-\mathcal{Z}) \biggr]  + 
\theta(\mathcal{Z}-l_h) \biggr\} \,  \; .
\label{eq:diffusion}
\end{equation}
Here $n$ is the number of elementary fields (2 for the pion, 3 for the
nucleon), $k_t = 0.350~\text{GeV/c}$ is the average transverse momentum of a
quark inside a hadron, $\mathcal{Z}$ is the distance the object has travelled
since its creation and $l_h \simeq 2p/\Delta M^2$ is the hadronic
expansion length, with $p$ the momentum of the final hadron and
$\Delta M^2$ the mass squared difference between the intermediate
prehadron and the final hadron state.  We adopted the values $\Delta M^2 = 1~
\text{(GeV)}^2$ for the proton and $\Delta M^2 = 0.7~\text{(GeV)}^2$
for the pion.

\begin{figure}[htb]
\centering
\includegraphics[width=8.6cm]{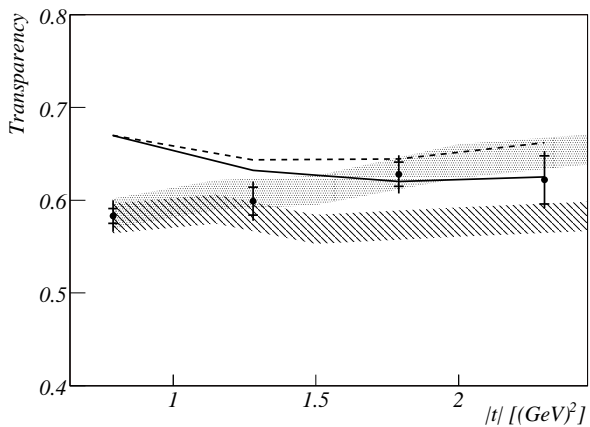}
\caption{The nuclear transparency extracted from
$^4\text{He}(\gamma,p\pi^-)$ versus the squared momentum transfer
$\mid t \mid$ at $\theta^\pi_{\text{c.m.}}=70^o$. The solid (dashed)
curve is the result of the RMSGA calculations without (with) color
transparency. The semi-classical model \cite{Gao:1996mg} results are
presented by the shaded areas: the hatched (dotted) area is a
calculation without (with) CT. Data from \cite{Dutta:2003mk}.}
\label{figHe70}
\end{figure}

\begin{figure}[htb]
\centering
\includegraphics[width=8.6cm]{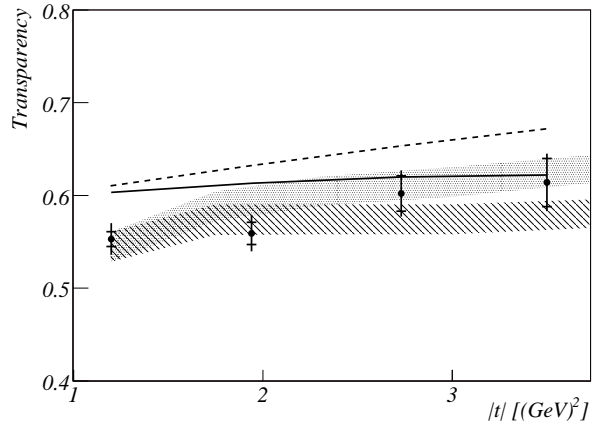}
\caption{As in Figure \ref{figHe70} but now for $\theta^\pi_{\text{c.m.}}=90^o$.}
\label{figHe90}
\end{figure}

In Figs. \ref{figHe70} and \ref{figHe90}, we present the results of
transparency calculations for $^4\text{He}$ together with the
experimental data and the predictions of the semi-classical model of
Ref.~\cite{Gao:1996mg}. In comparing transparency measurements with
theory, accurate modeling of the experimental cuts is required.
We adopt the following definition for the transparency
\begin{equation}\label{transp}
T=\frac{ \sum_i \sum_{\alpha} 
  Y(q_i)\left(\frac{d\sigma}{dE_{\pi_i} d\Omega_{\pi_i}
  d\Omega_{N_i}}\right)_{\text{RMSGA}}} {\sum_i \sum_{\alpha}
  Y(q_i) \left(\frac{d\sigma}{dE_{\pi_i} d\Omega_{\pi_i}
  d\Omega_{N_i}}\right)_{\text{RPWIA}}}\,,
\end{equation}
where $i$ denotes an event within the ranges set by the detector
acceptances and applied cuts. Further, $\sum _{\alpha} $ extends over
all occupied single-particle states in the target nucleus.  All cross
sections are computed in the lab frame.  Further, $Y(q)$ is the yield
of the reconstructed experimental photon beam spectrum for a certain
photon energy \cite{Dutta:2003mk}.  We assume that the elementary $
\gamma +n \rightarrow \pi ^- + p$ cross section $\frac { d \sigma
^{\gamma \pi}} {d \mid t \mid}$ in Eq.~(\ref{crossnoFSI}) remains
constant over the kinematical ranges which define a particular data
point. With this assumption the cross section $ \frac { d \sigma
^{\gamma \pi}} {d \mid t \mid} $ cancels out of the ratio
(\ref{transp}).  In order to reach convergence in the phase-space
averaging $\sum _i $ in Eq.~(\ref{transp}) we generated about one
thousand theoretical events within the kinematical ranges of the
experimental acceptances.  This was done for all data points, eight in
total, and corresponding kinematical ranges, of the Jefferson Lab
experiment.  Detailed kinematics for these data points can be found in
Ref.~\cite{Dutta:2003mk}.

The computed RMSGA nuclear transparencies are systematically about
10\% larger than the ones obtained in the semi-classical model. As can be
seen in Fig. \ref{figHe70}, our model predicts a rise in the transparency
for $\mid t \mid$ values below $1.2~\text{GeV}^2$.  This rise is due to
the minimum in the total proton-nucleon cross section in Eq.
(\ref{eq:profile}) for the proton momenta associated with these momentum
transfers.  
The RMSGA results overestimate the measured
transparencies at small $\mid t \mid$, but do reasonably well for the
higher values of $ \mid t \mid$.  Inclusion of CT effects tends to
increase the predicted transparency at a rate which depends on a
hard-scale parameter. Here, that role is played by the
momentum-transfer $\mid t \mid$.  Thus, inclusion of CT mechanisms
results in an increase of the nuclear transparency which grows with
the momentum transfer $ \mid t \mid$.  The magnitude of the increase
depends on the choice of the parameters in Eq.~(\ref{eq:diffusion}).
For the moment, there are no experimental constraints on their
magnitude.  As can be appreciated from Figs. \ref{figHe70} and
\ref{figHe90}, the RMSGA calculations predict comparable CT
effects as the semi-classical calculations.  We have to stress though
that the calculations with CT are normalized to the calculations
without CT for the data point with the lowest $|t|$ in the
semi-classical model.  We did not perform this normalization for our
calculations.  Our results without color transparency are in better
agreement with the experimental results than those with CT effects
included.  This is in disagreement with the semi-classical model whose
results with CT effects are in better agreement with the experimental
data.  We also have to point out that, although the calculations with CT
effects overestimate the experimental results for all data points, the
slope of this curve shows better agreement with the slope of the data than
the slope of the curves without CT effects.  
%Extra zin over waarom dit zo zou zijn...?

\begin{figure}[htb]
\centering
\includegraphics[width=8.5cm]{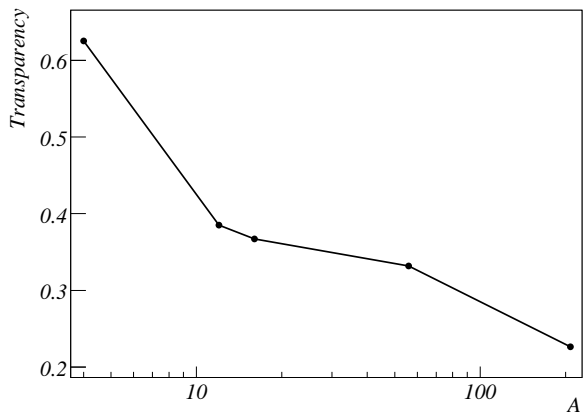}
\caption{A-dependence of the transparency.  Calculations were made for
$^4\text{He}$,$^{12}\text{C}$,$^{16}\text{O}$,$^{56}\text{Fe}$ and $^{208}\text{Pb}$
at $|t|=3.5~\text{GeV}^2$ for $\theta^\pi_{\text{c.m.}}=90^o$ without color transparency}
\label{figAdependence}
\end{figure}

To provide an idea of the A-dependence of the nuclear transparency
extracted from $A(\gamma,p \pi ^-)$, we plotted in
Fig.~\ref{figAdependence} the calculations for one data point for
several nuclei with the same kinematic cuts as before.  However, due
to these cuts, nucleon knockout from the innermost shells in the
heavier nuclei was not always possible for the generated events in the
calculations.  The transparency would even be lower for these heavier
nuclei if no cuts would be applied.

In summary, we have developed a quantum mechanical model based on a
relativistic extension to multiple-scattering Glauber theory to
calculate nuclear transparencies extracted from $A(\gamma,N \pi)$
processes.  The model can be applied to any even-even target nucleus
with a mass number $A \ge 4 $.  The nuclear transparency is the result
of the attenuating effect of the medium on the ejected proton and
pion, and is computed by means of a Glauber phase operator.  The
numerical computation of the latter, requires knowledge about $ \pi N
\rightarrow \pi N$ and $ N' N \rightarrow N' N$ cross sections, as
well as a set of relativistic mean-field wave functions for the
residual nucleus.  In contrast to alternative models, which adopt a
semi-classical approach, we treat FSI mechanisms at the amplitude
level in a quantum mechanical and relativistic manner. Comparison with
experimental results for helium shows no evidence of color
transparency in our model. Further progress will very much depend on
the availability of new data.  The model presented here can be readily
extended to electroproduction processes for comparison with the
forthcoming Jefferson Lab data.

We are grateful to D. Dutta for useful discussions.  This work is
supported by the Fund for Scientific Research Flanders and the
Research Board of Ghent University.  M. C. M. acknowledges
postdoctoral financial support from the Ministerio de Educaci\'{o}n y
Ciencia (MEC) of Spain.

\end{document}